\documentclass[
superscriptaddress,
 amsmath,amssymb,
 aps,
 prc,
 twocolumn,
]{revtex4-2}

\usepackage{graphicx}
\usepackage{dcolumn}
\usepackage{bm}
\usepackage{hyperref}
\usepackage[mathlines]{lineno}
\usepackage{xcolor}
\usepackage{soul}

\begin{document}


\title{Direct observation of $\beta$ and $\gamma$ decay from a high-spin long-lived isomer in $^{187}$Ta}
\author{J.~L.~Chen}
\affiliation{School of Physics, Beihang University, Beijing 100191, China}

\author{H.~Watanabe}
\email{hiroshi@buaa.edu.cn}
\affiliation{School of Physics, Beihang University, Beijing 100191, China}
\affiliation{RIKEN Nishina Center, 2-1 Hirosawa, Wako, Saitama 351-0198, Japan}
\affiliation{Wako Nuclear Science Center (WNSC), Institute of Particle and Nuclear Studies (IPNS), High Energy Accelerator Research Organization (KEK), Wako, Saitama 351-0198, Japan}

\author{P.~M.~Walker}
\affiliation{Department of Physics, University of Surrey, Guildford, Surrey GU2 7XH, United Kingdom}

\author{Y.~Hirayama}
\affiliation{Wako Nuclear Science Center (WNSC), Institute of Particle and Nuclear Studies (IPNS), High Energy Accelerator Research Organization (KEK), Wako, Saitama 351-0198, Japan}

\author{Y.~X.~Watanabe}
\affiliation{Wako Nuclear Science Center (WNSC), Institute of Particle and Nuclear Studies (IPNS), High Energy Accelerator Research Organization (KEK), Wako, Saitama 351-0198, Japan}

\author{M.~Mukai}
\affiliation{University of Tsukuba, Tsukuba, Ibaraki 305-0006, Japan}
\affiliation{RIKEN Nishina Center, 2-1 Hirosawa, Wako, Saitama 351-0198, Japan}
\affiliation{Wako Nuclear Science Center (WNSC), Institute of Particle and Nuclear Studies (IPNS), High Energy Accelerator Research Organization (KEK), Wako, Saitama 351-0198, Japan}

\author{C.~F.~Jiao}
\affiliation{School of Physics and Astronomy, Sun Yat-sen University, Zhuhai 519082, China}

\author{M.~Ahmed}
\affiliation{Wako Nuclear Science Center (WNSC), Institute of Particle and Nuclear Studies (IPNS), High Energy Accelerator Research Organization (KEK), Wako, Saitama 351-0198, Japan}
\affiliation{University of Tsukuba, Tsukuba, Ibaraki 305-0006, Japan} 

\author{M.~Brunet}
\affiliation{Department of Physics, University of Surrey, Guildford, Surrey GU2 7XH, United Kingdom}

\author{T.~Hashimoto}
\affiliation{Rare Isotope Science Project, Institute for Basic Science (IBS), Daejeon 305-811, Republic of Korea}

\author{S.~Ishizawa}
\affiliation{RIKEN Nishina Center, 2-1 Hirosawa, Wako, Saitama 351-0198, Japan}
\affiliation{Graduate School of Science and Engineering, Yamagata University, Yamagata 992-8510, Japan}
\affiliation{Wako Nuclear Science Center (WNSC), Institute of Particle and Nuclear Studies (IPNS), High Energy Accelerator Research Organization (KEK), Wako, Saitama 351-0198, Japan}

\author{F.~G.~Kondev}
\affiliation{Physics Division, Argonne National Laboratory, Lemont, Illinois 60439, USA}

\author{G.~J.~Lane}
\affiliation{Department of Nuclear Physics, RSPhys, Australian National University, Canberra, Australian Capital, Australia}

\author{Yu.~A.~Litvinov}
\affiliation{GSI Helmholtzzentrum für Schwerionenforschung, 64291 Darmstadt, Germany}

\author{H.~Miyatake}
\affiliation{University of Tsukuba, Tsukuba, Ibaraki 305-0006, Japan}
\affiliation{RIKEN Nishina Center, 2-1 Hirosawa, Wako, Saitama 351-0198, Japan}
\affiliation{Wako Nuclear Science Center (WNSC), Institute of Particle and Nuclear Studies (IPNS), High Energy Accelerator Research Organization (KEK), Wako, Saitama 351-0198, Japan}

\author{J.~Y.~Moon}
\affiliation{Rare Isotope Science Project, Institute for Basic Science (IBS), Daejeon 305-811, Republic of Korea}

\author{T.~Niwase}
\affiliation{RIKEN Nishina Center, 2-1 Hirosawa, Wako, Saitama 351-0198, Japan}
\affiliation{Wako Nuclear Science Center (WNSC), Institute of Particle and Nuclear Studies (IPNS), High Energy Accelerator Research Organization (KEK), Wako, Saitama 351-0198, Japan}
\affiliation{Department of Physics, Kyushu University, Nishi-ku, Fukuoka 819-0395, Japan}

\author{J.~H.~Park}
\affiliation{Rare Isotope Science Project, Institute for Basic Science (IBS), Daejeon 305-811, Republic of Korea}

\author{Zs.~Podolyák}
\affiliation{Department of Physics, University of Surrey, Guildford, Surrey GU2 7XH, United Kingdom}

\author{M.~Rosenbusch}
\affiliation{Wako Nuclear Science Center (WNSC), Institute of Particle and Nuclear Studies (IPNS), High Energy Accelerator Research Organization (KEK), Wako, Saitama 351-0198, Japan}

\author{P.~Schury}
\affiliation{Wako Nuclear Science Center (WNSC), Institute of Particle and Nuclear Studies (IPNS), High Energy Accelerator Research Organization (KEK), Wako, Saitama 351-0198, Japan}

\author{M.~Wada}
\affiliation{Wako Nuclear Science Center (WNSC), Institute of Particle and Nuclear Studies (IPNS), High Energy Accelerator Research Organization (KEK), Wako, Saitama 351-0198, Japan}
\affiliation{University of Tsukuba, Tsukuba, Ibaraki 305-0006, Japan}

\author{F.~R.~Xu}
\affiliation{School of Physics and State Key Laboratory of Nuclear Physics and Technology, Peking University, Beijing 100871, China}

\date{\today}

\begin{abstract}
$^{187}$Ta ($Z=73$, $N=114$) is located in the neutron-rich $A \approx 190$ region where a prolate-to-oblate shape transition via triaxial softness is predicted to take place. A preceding work on the $K^{\pi} = (25/2^-)$ isomer and a rotational band to which the isomer decays carried out by the same collaboration revealed that axial symmetry is slightly violated in this nucleus. This paper focuses on a higher-lying isomer, which was previously identified at 2933(14) keV by mass measurements with the Experimental Storage Ring at GSI. The isomer of interest has been populated by a multi-nucleon transfer reaction with a $^{136}$Xe primary beam incident on a natural tungsten target, using the KEK Isotope Separation System at RIKEN. New experimental findings obtained in the present paper include the internal and external $\beta$-decay branches from the high-spin isomer and a revised half-life of 136(24) s. The evaluated hindrances for $K$-forbidden transitions put constraints on the spin-parity assignment, which can be interpreted as being ascribed to a prolate shape with a five-quasiparticle configuration by model calculations.
\end{abstract}


\maketitle

\section{Introduction}
\label{sec:introduction}
The hafnium ($Z=72$), tungsten ($Z=74$), and osmium ($Z=76$) isotopes with $A \approx 190$ are located midway between the doubly mid-shell nucleus $^{170}_{~66}$Dy$_{104}$, which has an enlarged quadrupole (prolate) deformation at the ground state, and the doubly magic nucleus $^{208}_{~82}$Pb$_{126}$. Neutron-rich nuclei in this transitional region have attracted considerable interest both experimentally and theoretically, since different facets of nuclear shape manifest themselves at low excitation energies on account of a sensitive interplay between the microscopic [single-particle (SP)] and macroscopic (collective) degrees of freedom. For instance, in contrast to the large abundance of prolate-deformed shapes over the whole range of the nuclear chart \cite{PhysRevC.64.037301}, oblate states are found to be favored energetically in this specific mass region \cite{PhysRevC.72.047303,PhysRevC.77.064322,PhysRevC.79.031305}. Such a rare phenomenon takes place due to the occupation of low-$\Omega$, high-$J$ orbitals near the proton and neutron Fermi surfaces, which lie at the upper halves of the respective major shells, $Z = 50-82$ and $N = 82-126$, on the oblate side. In such shape-transition regions, where multiple energy minima coexist at prolate and oblate deformation in the potential-energy surface \cite{WOOD1992101}, the two different shapes can compete, and presumably interact, leading to the nuclear shape being soft with respect to the $\gamma$ (triaxial) degree of freedom, where $\gamma$ represents a deviation from axial symmetry of the ellipsoidal shape; $\gamma = 0^{\circ}$ and $60^{\circ}$ correspond to axially-symmetric prolate and oblate shapes, respectively, and $30^{\circ}$ for a maximally asymmetric nucleus that has three different radii in Cartesian coordinates. Understanding the behavior of such axially asymmetric nuclei near the ground state remains a challenge in nuclear structure studies. Furthermore, a global calculation based on the finite-range droplet macroscopic and microscopic models \cite{MOLLER20161} predicts substantial hexadecapole deformations ($\beta_4\lesssim$ -0.1) for nuclides in this region.  

Another interesting physics phenomenon observed in this mass region is the emergence of metastable nuclear excitations (isomers) at high spins \cite{PhysRevC.62.014301,Walker_2016}. The 16$^+$ state in $^{178}$Hf ($T_{1/2} = 31$ yr at an excitation energy of 2446 keV) \cite{PhysRevLett.82.695} is an exemplary case of a long-lived, highly excited isomer. As this nucleus has a well-deformed prolate shape, this isomerism is ascribed to the conservation of the $K$ quantum number, which is defined as the projection of the total nuclear spin on the symmetry axis of the deformed nucleus. However, this $K$-conservation law is often violated by various sources of symmetry breaking \cite{Dracoulis_2013}. Axially asymmetric deformation causes the mixing of configurations of different $K$ values, and thereby gives rise to a serious violation of $K$-forbidden transitions involved in radioactive decays,  as observed for the $J^{\pi} = K^{\pi} = (10^-)$ isomer in $^{192}$Os \cite{DRACOULIS2013330} and the (8$^+$) isomer in $^{190}$W \cite{PhysRevC.82.051304}. From an experimental point of view, neutron-rich nuclei in this transitional region can be produced by fragmentation of heavy-ion beams like $^{208}$Pb and $^{197}$Au \cite{PhysRevC.89.024616} or by multi-nucleon transfer (MNT) reactions with neutron-rich (stable) isotopes \cite{PhysRevLett.101.122701,PhysRevLett.115.172503}. Measurements of nuclear masses using the Experimental Storage Ring (ESR) at GSI have revealed the presence of long-lived ($T_{1/2} > 1$ s) isomers in the neutron-rich $Z = 72 - 76$ isotopes \cite{PhysRevLett.105.172501,PhysRevC.86.054321}.

The tantalum ($Z = 73$) isotopes intervene between the hafnium isotopes, which, as can be seen in potential-energy surfaces calculated on the $Q_0- \gamma$ plane \cite{Robledo_2009}, undergo a transition from a prolate shape to an oblate one within neutron numbers of $114-118$, and the tungsten isotopes that exhibit a gradual shape transition via $\gamma$ softness in a much broader range of neutron number. We draw particular attention to $^{187}$Ta$_{114}$ approaching the critical-point neutron number $N = 116$ of the predicted prolate-to-oblate shape transition for the lower-$Z$ nuclides and of the maximum axial asymmetry ($\gamma \approx 30^{\circ}$) for the higher-$Z$ nuclides. Using high mass resolving power at the ESR, the ground state and two long-lived isomers in $^{187}$Ta (hereinafter denoted as $^{187}$Ta$^g$ and $^{187}$Ta$^{m1, m2}$) were identified with their masses, which were translated into excitation energies of 1789(13) keV ($^{187}$Ta$^{m1}$) and 2933(14) keV ($^{187}$Ta$^{m2}$) \cite{PhysRevLett.105.172501,PhysRevC.86.054321}. More recently, our previous works reported on the internal decay from $^{187}$Ta$^{m1}$ \cite{PhysRevLett.125.192505}, as well as the ground-state $\beta$ decays~\cite{PhysRevC.105.034331}, and revised the values of the neutral-atom $^{187}$Ta$^{m1}$ half-life and excitation energy to 7.3(9) s and 1778(1) keV, respectively, from those measured in the form of bare ions at the ESR. $^{187}$Ta$^{m1}$ was assigned $I^{\pi} = K^{\pi} = (25/2^-$) based on the observation of decay patterns towards the rotational band built on the 9/2$^-$[514] single-proton state. It turned out that axial symmetry is slightly broken from the observed signature splitting in the 9/2$^-$ rotational band and the reduced hindrance for the $E2$ decay of the ($25/2^-$) isomer in comparison with the neighboring isotopes/isotones \cite{PhysRevC.80.024321,REED2016311}. However, the decay properties of a higher-lying isomer, $^{187}$Ta$^{m2}$, and its characterization remain to be scrutinized. This paper is dedicated to the first direct observation of decay radiations from $^{187}$Ta$^{m2}$ in order to bring a series of works concerning $^{187}$Ta at the KEK Isotope Separation System (KISS) facility~\cite{PhysRevLett.125.192505,PhysRevC.105.034331} to completion.

\section{Experimental details}
\label{sec:formalism}
The experiment was performed using the KISS setup at the RIKEN Nishina Center in Japan. A 50-pnA primary beam of $^{136}$Xe was accelerated up to 7.2 MeV per nucleon using the RIKEN Ring Cyclotron and incident on a natural tungsten target (28\%{} $^{186}$W) with a thickness of 5 $\mu$m mounted on a rotating wheel. Following MNT reactions between the beam and target nuclides, the reaction products were stopped in the gas cell filled with high-pressure  (80 kPa) gaseous argon and subsequently transported by a gas flow towards the cell outlet where a laser ionization technique is applied for element selectivity. The $^{187}$Ta$^{+}$ ions were extracted at an energy of 20 keV by the rf ion guides, followed by mass separation through the dipole magnet with a mass resolving power $A/{\Delta A} =900$.  For more details about the technical development and recent progress of the KISS equipment, the reader is referred to Refs.~\cite{HIRAYAMA20154,HIRAYAMA201652,HIRAYAMA201711,HIRAYAMA_2024}.

During five days of data taking, the $^{187}$Ta$^{+}$ ions were extracted with an average intensity of 1.5 ions/s and transported either to a decay spectroscopy setup \cite{MUKAI20181} or to a multi-reflection time-of-flight (MRTOF) device for high-resolution mass measurements \cite{Moon_RIKEN}, the latter of which was installed at $90^{\circ}$ with respect to the original KISS beam axis. The beam trajectory was switched between the decay station and MRTOF using an electrostatic deflector located about 1 m upstream of the terminal of the beamline. For decay measurements, the secondary beams were distributed to the decay station and MRTOF during given periods of beam-on ($T_{\text{on}}$) and beam-off ($T_{\text{off}}$), respectively, in one cycle. The adopted $T_{\text{on}}$/$T_{\text{off}}$ conditions in this experiment were 30/30, 300/300, and 1800/1800 s in order to accommodate decay half-lives $^{187}$Ta$^{g, m1, m2}$ reported previously at the ESR mass measurements \cite{PhysRevLett.105.172501,PhysRevC.86.054321}. A total of about 1.36$\times$ 10$^5$ $^{187}$Ta nuclides were collected with $T_{\text{on}}$/$T_{\text{off}}$ = 1800/1800 s and used for the data analysis described in Sec. $\text{\uppercase\expandafter{\romannumeral3}}
$.  

The decay station consists of a tape-transport system with a 12-$\mu$m-thick aluminized Mylar tape, on which the 20-keV secondary beams were implanted, a multi-segmented proportional gas counter (MSPGC) \cite{MUKAI2020421,HIRAYAMA2021165152}, and four large-volume Clover-type HPGe detectors arranged in a close geometry. A full-energy peak efficiency of 7.8\% (15\%) was achieved for a $\gamma$ ray at 1 MeV (150 keV) with the four Clover-type HPGe detectors operated in the so-called add-back mode. The residual radioactivity was removed from the implantation position by rolling up the Mylar tape by about 30 cm at the end of each cycle. The MSPGC covered 80\% of a 4$\pi$ solid angle with two concentric layers of 16 gas counters around the Mylar tape. The MSPGC has a capability to distinguish between high-energy $\beta$ rays, which can penetrate adjacent gas counters in the inner and outer layers, and low-energy ($\lesssim 100$ keV) internal-conversion electrons (ICEs) that are likely to be detected only by the inner layer.


\begin{figure*}[htp]
    \centering
    \includegraphics[width=1\textwidth]{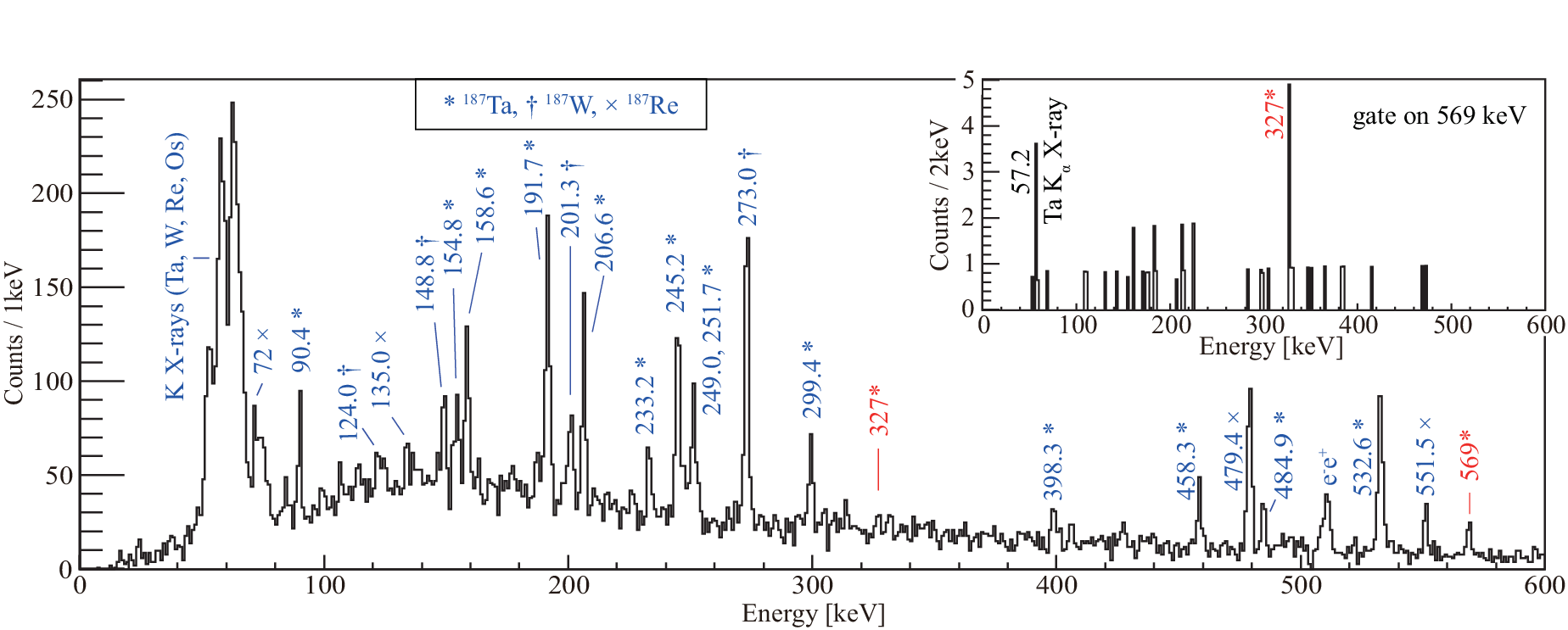}
\caption{Gamma-ray energy spectrum measured in coincidence with MSPGC(M = 1) within an electron-$\gamma$ time window of $\pm$200~ns after implantation of $^{187}$Ta. Transitions emitted in the $^{187}$Ta$^{m1}$ $\rightarrow$ $^{187}$Ta$^g$, $^{187}$Ta$^{g}$ $\rightarrow$ $^{187}$W, and $^{187}$W $\rightarrow$ $^{187}$Re decays are marked with asterisks, daggers and crosses, respectively. Previously unreported $\gamma$ rays are colored in red. The inset shows a $\gamma$-ray energy spectrum measured by gating on the 569-keV $\gamma$ ray.}
    \label{fig:1}
\end{figure*}  

\section{Results of data analysis}
\label{sec:formalism}

The following analysis of $\gamma$-ray spectra always uses coincidences with the MSPGC: The analysis method that requires events only in the inner layer is referred to as the multiplicity-1 (M = 1) mode, while it is denoted as the multiplicity-2 (M = 2) mode when a pair of the inner- and outer-layer gas counters is fired. 

Figure~\ref{fig:1} shows the $\gamma$-ray energy spectrum measured in coincidence with MSPGC(M = 1) within an electron-$\gamma$ time window of $\pm$200 ns. All of the labeled $\gamma$-ray and X-ray lines, except peaks at 327- and 569-keV, have been identified in our previous works for the decay of $^{187}$Ta$^{g}$~\cite{PhysRevC.105.034331} and $^{187}$Ta$^{m1}$ \cite{PhysRevLett.125.192505}. It should be noted that the 569-keV $\gamma$ ray could have been clearly observed neither in coincidence with MSPGC(M = 2) nor with gates on any $\gamma$-ray peaks corresponding to the transitions between $^{187}$Ta$^{m1}$ and $^{187}$Ta$^{g}$ (see Fig.~1 in Ref.~\cite{PhysRevC.105.034331} and Fig. 3 in Ref.~\cite{PhysRevLett.125.192505}), implying that the emission of the 569-keV $\gamma$ ray takes place following internal decay of a long-lived isomer other than $^{187}$Ta$^{m1}$. The most probable candidate for the origin of the 569-keV transition is $^{187}$Ta$^{m2}$, which was previously identified in the ESR measurements \cite{PhysRevLett.105.172501,PhysRevC.86.054321}. This assignment can be supported by the fact that the Ta K$_{\alpha}$-X rays have been observed in coincidence with the 569-keV $\gamma$ ray, as exhibited in the inset of Fig.~\ref{fig:1}. It was also found that the $\gamma$ rays of 569 and 327 keV are in mutual coincidence. Thus, the 569-327-keV cascade is suggested to feed $^{187}$Ta$^{m1}$ in the decay from $^{187}$Ta$^{m2}$, though their order is ambiguous. The energy sum for these two transitions is 259 keV smaller than the gap in mean energy between $^{187}$Ta$^{m1}$ (1778(1) keV \cite{PhysRevLett.125.192505}) and $^{187}$Ta$^{m2}$ (2933(14) keV \cite{PhysRevLett.105.172501}), and this residual energy is much larger than the square root of the sum of squares of the uncertainties. However, no $\gamma$-ray peaks are visible around 260 keV or lower with a gate on the 569-keV $\gamma$ ray (see the inset of Fig.~\ref{fig:1}), presumably because the long-lived $^{187}$Ta$^{m2}$ undergoes internal decay through a high-multipolarity transition dominated by internal conversion processes. As such, the emitted ICEs could be detected by MSPGC and served for coincidence measurements with the subsequent $\gamma$ rays.

\begin{figure}[tb!]
    \centering
    \includegraphics[width=0.49\textwidth]{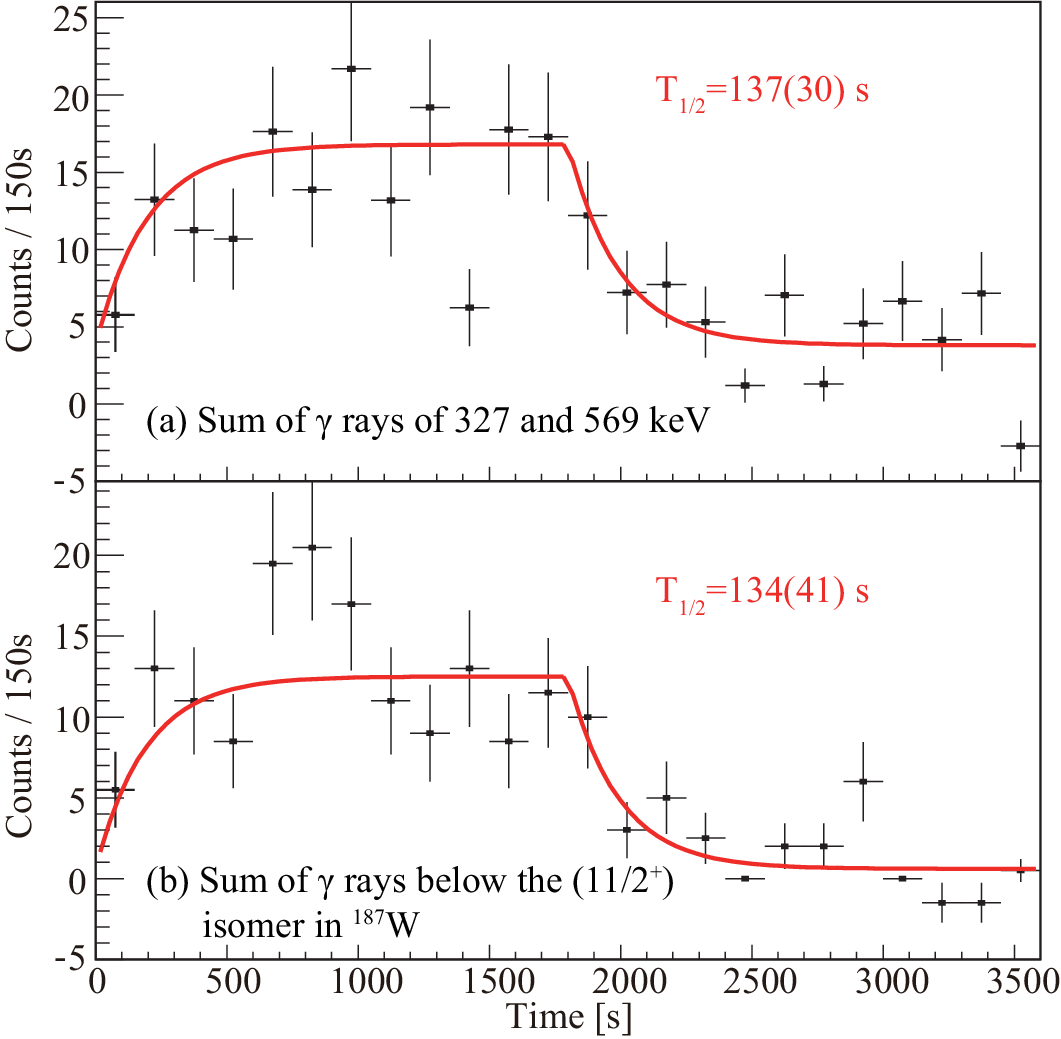}
\caption{Background-subtracted $\gamma$-ray time distributions measured with a sum of gates on (a) the 327- and 569-keV transitions in coincidence with MSPGC(M = 1) and (b) those below the ($11/2^+$) isomer in $^{187}$W in coincidence with MSPGC(M = 2) within an electron-$\gamma$ time window of 0.2-0.8 $\mu$s. The measurements of the growth and decay curves were carried out using macroscopically pulsed $^{187}$Ta$^+$ beams with a $T_{\text{on}}/T_{\text{off}}$ condition of 1800/1800 s. The red solid lines represent the result of a log-likelihood fit.}
    \label{fig:2}
\end{figure}

Figure~\ref{fig:2}(a) shows a time distribution measured in the $T_{\text{on}}/T_{\text{off}}$ = 1800/1800 s dataset with a sum of gates on the 569- and 327-keV $\gamma$ transitions with appropriate background subtraction. A log-likelihood fit to the time distribution by taking into account the growth and decay curves with a uniform background yields a half-life of 137(30) s, which differs appreciably from those of $^{187}$Ta$^{g}$ (283(10) s \cite{PhysRevC.105.034331}) and $^{187}$Ta$^{m1}$ (7.3(9) s \cite{PhysRevLett.125.192505}). 

\begin{figure}[htb!]
    \centering
    \includegraphics[width=0.49\textwidth]{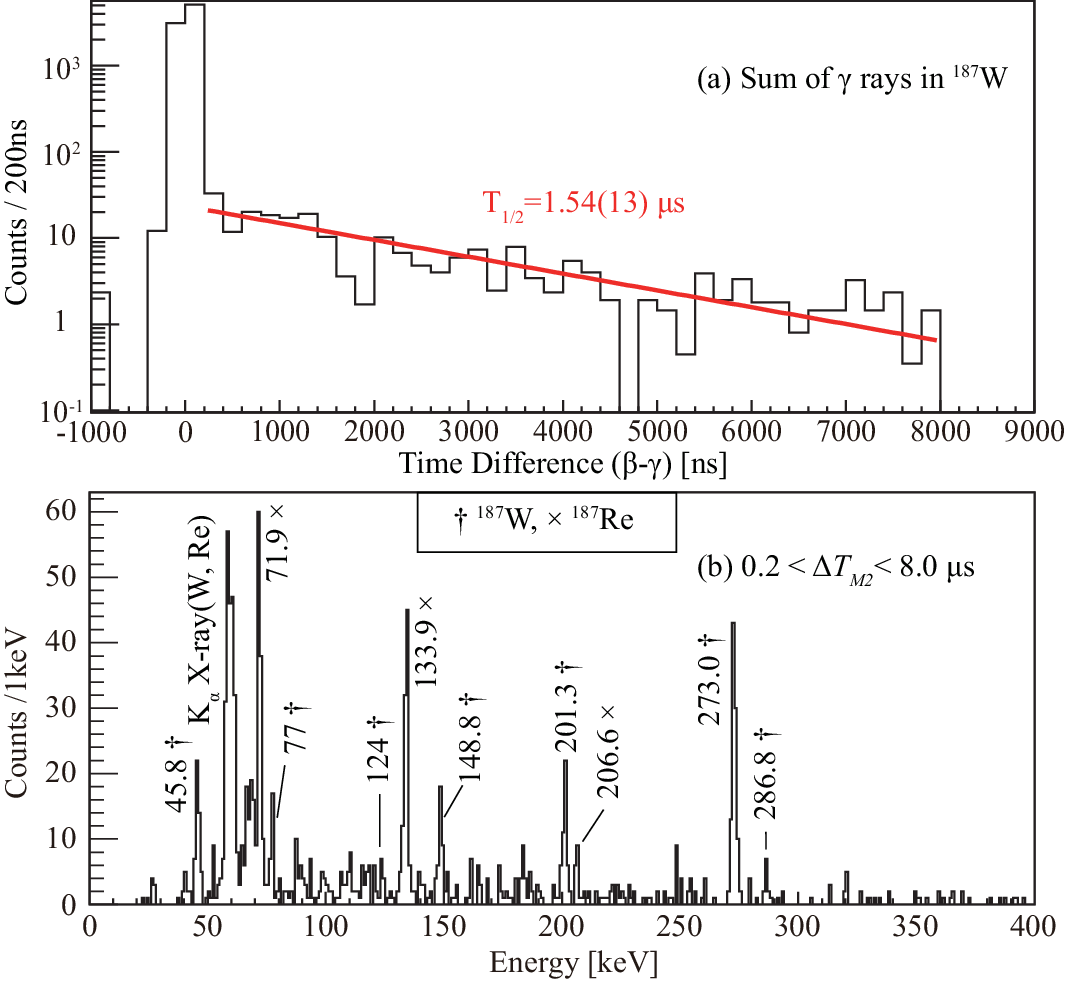}
\caption{(a) Electron-$\gamma$ time differential spectrum measured with a sum of gates on the $\gamma$ rays below the ($11/2^+$) state and an associated fit to the time slope. (b) $\gamma$-ray energy spectrum measured in coincidence with MSPGC(M = 2) within a time window of 0.2-8.0 $\mu$s. Transitions in $^{187}$W and $^{187}$Re are marked with daggers and crosses, respectively.}
    \label{fig:bd}
\end{figure}

\begin{figure}[htp!]
    \centering
    \includegraphics[width=0.49\textwidth]{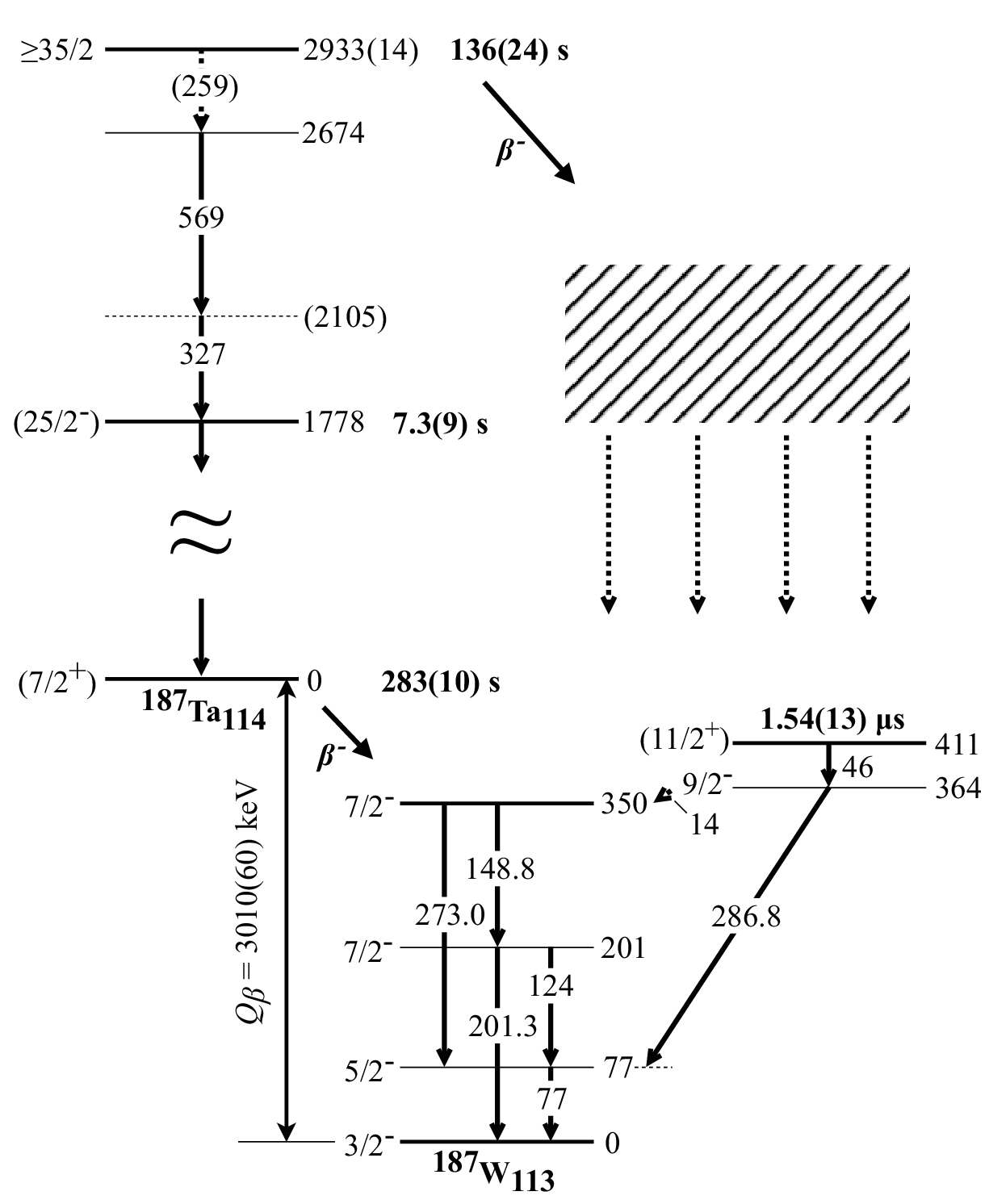}
\caption{Proposed decay scheme of $^{187}$Ta and levels in the daughter nucleus $^{187}$W. The numbers in bold letters represent the level half-lives determined in Refs.~\cite{PhysRevLett.125.192505} and\cite{PhysRevC.105.034331} and the present paper. The hatched area indicates the presence of intermediate states in $^{187}$W, which are populated by the $\beta$ decay from the 2933(14)-keV isomer in $^{187}$Ta and subsequently decay towards the lower-lying levels. The dotted arrows represent presumed $\gamma$ transitions. Spins, parities, and excitation/$\gamma$-ray energies in parentheses denote tentative assignments. The levels and transitions between the  ($25/2^-$) isomer and the ground state in $^{187}$Ta, which are shown in Fig.~2 in Ref.~\cite{PhysRevLett.125.192505}, are omitted. The Q$_{\beta}$ value for the $^{187}$Ta$\rightarrow$$^{187}$W $\beta$ decay is taken from Ref.~\cite{Wang_2021}.}
    \label{fig:3}
\end{figure}

The possibility of $\beta$ decay from $^{187}$Ta$^{m2}$ was suggested in the ESR experiment~\cite{PhysRevLett.105.172501,PhysRevC.86.054321}. We therefore explored $\beta$-decay branches from $^{187}$Ta$^{m2}$ towards the daughter nucleus $^{187}$W. In Fig.~\ref{fig:bd}(a), a sum of electron-$\gamma$ time differential spectra for transitions in $^{187}$W \cite{PhysRevC.105.034331,PhysRevC.71.067301} measured with MSPGC(M = 2) exhibits a long tail at late times. A fit to the delayed component results in a half-life of 1.54(13) $\mu$s, consistent with the reference value for the (11/2$^+$) isomeric state at 411 keV in $^{187}$W \cite{PhysRevC.71.067301}. The 46-keV $\gamma$ ray, which deexcites the (11/2$^+$) isomer~\cite{PhysRevC.71.067301}, can be unambiguously observed by choosing an electron-$\gamma$ time window of 0.2 - 8.0 $\mu$s, as shown in Fig.~\ref{fig:bd}(b). However, it is unlikely that the ($11/2^+$) state is populated directly from the (7/2$^+$) state of $^{187}$Ta$^{g}$ via a second-forbidden $\beta$ decay in competition with strong first-forbidden transitions towards the lower-lying negative-parity states in $^{187}$W. Furthermore, no transitions have been assigned as feeding the ($11/2^+$) isomer at 411 keV in the decay of $^{187}$Ta$^{g}$~\cite{PhysRevC.105.034331} despite the sufficiently large $Q_{\beta}$ value of 3010(60) keV~~\cite{Wang_2021}. The value of the half-live derived from a fit to the summed gated time spectra for the delayed $\gamma$ rays below the (11/2$^+$) isomeric state in $^{187}$W is in reasonable agreement with that for the internal-decay transitions from $^{187}$Ta$^{m2}$, as compared in Figs.~\ref{fig:2}(a) and \ref{fig:2}(b). Thus, it is rational to conclude that there is a $\beta$-decay pathway from $^{187}$Ta$^{m2}$ towards higher-spin states, which subsequently populate the (11/2$^+$) isomeric state in $^{187}$W. Several transitions are known to feed the (11/2$^+$) isomeric state in $^{187}$W \cite{PhysRevC.77.047303}. In the present paper, however, no such feeding transitions have been clearly observed as discrete $\gamma$-ray lines despite a careful analysis of $\gamma$-$\gamma$ coincidences with respect to the delayed transitions below the (11/2$^+$) isomer. A possible reason for their non-observation is that the $^{187}$Ta$^{m2}$ $\beta$ decay disperses into some states at high spin in the daughter nucleus $^{187}$W, making it difficult to identify the respective $\gamma$ rays with the limited statistics. Identification of the intermediate states populated by the $\beta$ decays of $^{187}$Ta$^{m2}$ remains a challenge for future experiments.

The half-life of $^{187}$Ta$^{m2}$ was determined to be 136(24)~s by taking a weighted average of the half-lives extracted from the time distributions for the internal and external decay branches shown in Figs.~\ref{fig:2}(a) and \ref{fig:2}(b), respectively. This neutral-atom half-life measured at rest is much shorter than the lower limit ($>$5 min) obtained for fully stripped ions in the ESR mass measurements~\cite{PhysRevLett.105.172501,PhysRevC.86.054321}, due partly to the fact that internal-conversion processes can not occur in bare ions during flight through the ESR. A more precise value of the bare-ion half-life, which should be measured with an optimum ESR setting~\cite{PhysRevC.86.054321}, is needed to put a constraint on the total conversion coefficient  ($\alpha_T$) for a transition de-exciting $^{187}$Ta$^{m2}$ in comparison with the the neutral-atom half-life measured in the present paper. 

Figure~\ref{fig:3} displays the proposed decay scheme of $^{187}$Ta$^{m2}$. As already mentioned, the order of the 569- and 327-keV transitions was undetermined and the 259-keV transition, which should be highly converted, was not observed. It is to be noted that, if either of the 569- or 327-keV transitions deexcited $^{187}$Ta$^{m2}$, a subsequent level that would decay via a high-mulitpolarity 259-keV transition should be long lived: This is at variance with the fact that the 569- and 327-keV $\gamma$ rays were observed in prompt coincidence with MSPGC(M = 1). Thus, in the following discussion, the internal transition (IT) branch is assumed to proceed only with the unobserved 259-keV transition. The total decay constant of $^{187}$Ta$^{m2}$, $\lambda_{m2} = \frac{0.693}{136(24)} = 5.1(9)\times10^{-3}$ s$^{-1}$, is the sum of the partial decay rates for the IT ($\lambda_{\text{IT}}$) and $\beta$ decay ($\lambda_{\beta}$), which can be expressed as
\begin{equation}\label{lambdam2}
\begin{split}
\lambda_{m2} = (1 + \alpha_{T, 259})\times\lambda_{\gamma, 259} + \lambda_{\beta} \\
= (1 + \alpha_{T, 259} + R)\times\lambda_{\gamma, 259},
\end{split}
\end{equation}
where $\alpha_{T, 259}$ and $\lambda_{\gamma, 259}$ are the total conversion coefficient and the partial decay rate for $\gamma$-ray emission of the 259-keV transition, respectively, and $R = \frac{\lambda_{\beta}}{\lambda_{\gamma, 259}}$ represents the ratio of partial decay rates for $\beta$ and $\gamma$ branches from $^{187}$Ta$^{m2}$. The values of $\alpha_{T, 259}$ can be calculated by the BrIcc code \cite{KIBEDI2008202} for possible multipolarities $L_{259}$, as given in the second column in Table~\ref{tab:1}. To evaluate $\lambda_{\gamma, 259}$ and the corresponding transition probability for the IT-$\gamma$ branch, it is necessary to estimate the $\beta / \gamma$ branching ratio $R$.

The value of $R$ can be evaluated as follows: The electron-$\gamma$ coincidence measurement with MSPGC(M = 1), shown in Fig.~\ref{fig:1}, yielded 41(10) events of the 569-keV $\gamma$ ray, which can be associated with the total yield of $^{187}$Ta$^{m2}$ ($Y_{m2}$) as
\begin{equation}\label{eq2}
\begin{split}
{N_{\gamma, 569}}&={Y_{m2}}\times{p_{\gamma, 259}}\times\frac{1 + \alpha_{T, 259}}{1 + \alpha_{T, 569}}\times\varepsilon_{\rm{Ge(569)}}\\
&\times\left(\frac{\alpha_{T, 259}}{1 + \alpha_{T, 259}} + \frac{\alpha_{T, 327}}{1 + \alpha_{T, 327}}\right)\times\varepsilon_{\rm{GC(M=1)}},
\end{split}
\end{equation}
where, $p_{\gamma, 259}$ denotes the proportion of the IT branch through emission of the 259-keV $\gamma$ ray. Meanwhile, the number of observed counts of the 46-keV $\gamma$ ray in the daughter nucleus $^{187}$W reflects $p_{\beta}$, the proportion of the $\beta$-decay branch that subsequently feeds the (11/2$^+$) isomer, thus, 
\begin{equation}\label{eq3}
\begin{split}
N_{\gamma, 46} = Y_{m2}\times{p_{\beta}}\times\varepsilon_{\rm{GC(M=2)}}\times\frac{1}{1 + \alpha_{T, 46}}\times\varepsilon_{\rm{Ge(46)}}.
\end{split}
\end{equation}
Note that, since there may be pathways that bypass the (11/2$^+$) isomer in $^{187}$W, the value of $p_{\beta}$  should be considered as a lower limit of the $\beta$-decay branch from $^{187}$Ta$^{m2}$, therefore, $R>\frac{p_{\beta}}{p_{\gamma,259}}$. 

In Eqs.~(\ref{eq2}) and (\ref{eq3}), the energy dependent $\gamma$-ray (add-backed) full-energy peak efficiency $\varepsilon_{{\rm Ge}(E_{\gamma})}$ was estimated with $^{152}$Eu and $^{133}$Ba standard sources placed at the beam implantation position. For the detection efficiency of the MSPGC, $\varepsilon_{\rm{GC(M=2)}} = 58(5)$~\% was adopted from Fig.~15 in Ref.~\cite{MUKAI20181} with 5.53(6) MeV, which corresponds to the difference in energy between $^{187}$Ta$^{m2}$ and the $(11/2^+)$ isomer in $^{187}$W. The value of $\varepsilon_{\rm{GC(M=1)}}$ was estimated to be 43(8)~\% using the intensity of $\gamma$ rays from the $(9^+)$ isomeric state in $^{186}$Ta measured in the same dataset~\cite{PhysRevC.104.024330}. 

Concerning Eq.~(\ref{eq3}), the efficiency-corrected total transition intensity for the 46-keV $\gamma$ transition, $(1+{\alpha}_{T,46})\times\frac{N_{\gamma,46}}{\varepsilon_{\rm{Ge(46)}}}$, is balanced with the sum of the transitions of 149, 273, and 287 keV in $^{187}$W (see Fig.~\ref{fig:3}), where these $\gamma$ rays are measured in electron-$\gamma$ delayed coincidence within the same time window, e.g. $0.2-8$ $\mu$s, as demonstrated in Fig.~\ref{fig:bd}(b). As such, a weighted average of the total transition intensities was taken between the 46-keV transition and the sum of the subsequent three transitions for the evaluation of $p_{\beta}$. The loss of events that occur out of the applied electron-$\gamma$ time window was also taken into account. 

\begin{table*}[htp!]
\caption{Properties of the presumed 259-keV de-excitation of $^{187}$Ta$^{m2}$. See the text for details.}
\label{tab:1}
\begin{ruledtabular}
\begin{tabular}{cccccc}

$L_{259}$ & $\alpha_{T, 259}$ & $p_{\beta}/p_{\gamma, 259}$ & $\lambda_{\gamma, 259}$ [s$^{-1}$] & $B$($L_{259}$) [W.u.] & $\log F$ \\
 \hline
 $E1$ &  $0.03$ & $0.10(3) \sim 0.35(12)$ & $\{3.7(7) \sim 4.5(8)\} \times 10^{-3}$ & $\{6.3(12) \sim 7.7(14)\} \times10^{-17}$ & $16.11(8) \sim 16.20(9)$  \\
 $M1$ &  $0.30$ & $0.6(2) \sim 1.0(3)$ & $\{2.2(5) \sim 2.6(6)\}\times 10^{-3}$  & $\{4.1(9) \sim 4.8(10)\} \times 10^{-15}$ & $14.32(9) \sim 14.39(10)$  \\ 
 $E2$ &  $0.13$ & $0.29(10) \sim 0.58(19)$ & $\{3.0\pm0.6 \sim 3.6\pm0.7\} \times 10^{-3}$ & $\{3.3(7) \sim 4.0(8)\} \times 10^{-11}$ & $10.40(8) \sim 10.48(9)$ \\
 $M2$ &  $1.32$ & $2.7(9) \sim 3.3(11)$  & $\{0.9(2) \sim 1.0(3)\} \times 10^{-3}$ & $\{1.1(3) \sim 1.2(3)\} \times 10^{-9}$ & $8.92(11) \sim 8.97(12)$  \\
 $E3$ &  $0.69$ & $1.4(5) \sim 1.9(6)$ & $\{1.4(4) \sim 1.6(4)\} \times 10^{-3}$ & $\{1.6(4) \sim 1.8(4)\} \times 10^{-5}$ & $4.75(10) \sim 4.81(11)$ \\
 $M3$ &  $5.05$ & $10(3) \sim 12(4)$ & $\{2.8(8) \sim 3.2(9)\} \times 10^{-4}$ & $\{3.3(9) \sim 3.6(10)\} \times 10^{-4}$  & $3.44(12) \sim 3.49(12)$  \\
 $E4$ &  $4.16$ & $8(3) \sim 10(3)$ & $\{3.4(10) \sim 3.8(10)\} \times 10^{-4}$ & $5.3(15) \sim 5.9(16)$  & $-0.77(12) \sim -0.72(12)$ \\
 $M4$ &  $20.7$ & $42(14) \sim 48(16)$ & $\{7(2) \sim 8(2)\} \times 10^{-5}$ & $\{1.2(4) \sim 1.3(4)\} \times 10^2$ & $-2.13(12) \sim -2.08(13)$ \\
\end{tabular}
\end{ruledtabular}
\end{table*}

The value of $p_{\gamma,259}$, which can be derived from $N_{\gamma,569}$ in Eq.~(\ref{eq2}), varies with $\alpha_T$ of the 259-, 327-, and 569-keV transitions. Here, the 327- and 569-keV $\gamma$ rays are considered of either $E1$, $M1$, or $E2$ character, \textit{i.e}., for a given multipolarity $L_{259}$ (and $\alpha_T$) of the 259-keV transition, there are $3 \times 3$ combinations of $\alpha_{T,327}$ and $\alpha_{T,569}$. (It is unlikely that the 327- and 569-keV transitons have an $M2$ or higher multipolarity since they were observed in electron-$\gamma$ and $\gamma$-$\gamma$ coincidence within a prompt time window.) Among the possible nine values of $p_{\gamma,259}$, it was found that the largest (smallest) one is given when the 327- and 569-keV transitions have $E1$ ($M1$) and $M1$ ($E1$) multipolarities, respectively, in every case of $L_{259}$ considered here. 
For each $L_{259}$, the range of $\frac{p_{\beta}}{p_{\gamma,259}}$ calculated with 1$\sigma$ confidence limits for the minimum and maximum values is presented in the third column in Table~\ref{tab:1}. 
Since the quantity $\frac{p_{\beta}}{p_{\gamma,259}}$ puts a lower limit on the $\beta / \gamma$ branching ratio $R$, the values of $\lambda_{\gamma,259}$ and the corresponding reduced transition probabilities $B(L_{259})$ given in the fourth and fifth columns in Table~\ref{tab:1} must be considered as the upper limits.

\section{Discussion}
\label{sec:discussion}
Excited state lifetimes become longer when electromagnetic transitions proceed with low energy and/or high multipolarity, as well as when a large change in the $K$ quantum number is involved in well-deformed nuclei. In the latter case, it is empirically known that the hindrance, which is defined as the ratio of the partial $\gamma$-ray emission lifetime relative to the Weisskopf single-particle estimate, $F = \frac{\tau_{\gamma}}{\tau_W}$, rises exponentially with the difference in $K$ between the initial and final states (or, the shortfall with respect to the transition multipolarity $L$, known as the degree of $K$ forbiddenness, $\nu = \mathit{\Delta}K - L$), and the so-called reduced hindrance $f_\nu = F^\frac{1}{\nu}$ ranges typically from 30 to 200. Following the previous suggestion by L{\"o}bner~\cite{LOBNER1968369} about the observed hindrances that would have an $L$-dependent offset, Kondev \textit{et~al.}~\cite{KONDEV201550} have adopted an approach to separate the hindrance into an intrinsic hindrance ($F_0$) that depends on $L$ and a factor ($f_0$) to the $\nu$-th power: The product of these two terms, i.e. $F = F_0 \times f_0^\nu$, or more conveniently, its logarithmic form, ${\rm log}F = {\rm log}F_0 + \nu \times {\rm log}f_0$, was used to fit the available data of hindrances as a function of $\mathit{\Delta}K$~\cite{KONDEV201550}. For $E1$, $M1$, and $E2$ multipolarities, the values of $F_0$ and $f_0$ were evaluated from a linear fit to the centroids of the ${\rm log}F$ distribution, whereas the individual data were analyzed for $E3$ (see Figs. 15 and 16, and Table C in Ref.~\cite{KONDEV201550}). There is no evaluation for other multipolarities, but it is possible to analyze the $M2$ hindrances in a similar way to the $E3$ cases.  

We considered what value of the $K$ quantum number $^{187}$Ta$^{m2}$ can possess based on the systematics of the ${\rm log}F$ values. The sixth column in Table~\ref{tab:1} shows the ranges of ${\rm log}F$ evaluated for the respective $L_{259}$. If the 259-keV transition were of $E1$, $M1$, or $E2$ character, the ${\rm log}F$ value would be quite large ($\textgreater$10), corresponding to $\mathit{\Delta}K$ larger than 10$\hbar$ in comparison to Fig.~15 in Ref.~\cite{KONDEV201550}. Such low-$L$ and large-$\mathit{\Delta}K$ transitions, if any, should be followed by a cascade of four or more $\gamma$ rays within the rotational band built on a lower-$K$ state. This is inconsistent with what was observed in the present paper, whereby the possibility of $E1$, $M1$, and $E2$ de-excitation from $^{187}$Ta$^{m2}$ can be ruled out. On the other hand, an $E3$ decay with ${\rm log}F = 4.7 \sim 4.9$ is expected to have a $\mathit{\Delta}K$ of 5$\hbar$ or $6 \hbar$ by considering the parameters $F_0$ and $f_0$ with their uncertainties determined for $E3$ hindrances in Ref.~\cite{KONDEV201550}. 
A similar analysis of the individual $M2$ hindrances compiled in Ref.~\cite{KONDEV201550} implies that an $M2$ transition with ${\rm log}F = 8.8 \sim 9.1$ can proceed with $\mathit{\Delta}K$ = 6$\hbar$ or 7$\hbar$. 
For $M3$, there are few data available to see the trend of ${\rm log}F$ with different values of $\mathit{\Delta}K$. Given $\mathit{\Delta}K$ = 5$\hbar$ and 6$\hbar$ for an $M3$ transition, ${\rm log}F = 3.3 \sim 3.6$ results in $f_{\nu} = 45 \sim 63$ and $13 \sim 16$, respectively, comparable to $f_{\nu} = 27(1)$ obtained for the $E2$ decay of $^{187}$Ta$^{m1}$~\cite{PhysRevLett.125.192505}, indicating significant $K$ mixing due to a loss of axial symmetry. The cases of $E4$ and $M4$ are unlikely to take place as a pure multipolarity since they would be enhanced (${\rm log}F < 0$). In conclusion, plausible candidates for the $^{187}$Ta$^{m2}$ de-excitation are of $M2$ ($\mathit{\Delta}K$ = 6$\hbar$ or 7$\hbar$), $E3$ ($\mathit{\Delta}K$ = 5$\hbar$ or 6$\hbar$), and $M3$ ($\mathit{\Delta}K$ = 5$\hbar$ or 6$\hbar$) character. The $K$ quantum number of the (observed) state to which $^{187}$Ta$^{m2}$ decays is considered to be equal to or larger than $^{187}$Ta$^{m1}$ ($K=25/2$). Thus, $^{187}$Ta$^{m2}$ is expected to have $K \geq 35/2$ from the above argument on the possible multipolarity.

\begin{figure}[t]
    \centering
    \includegraphics[width=0.5\textwidth]{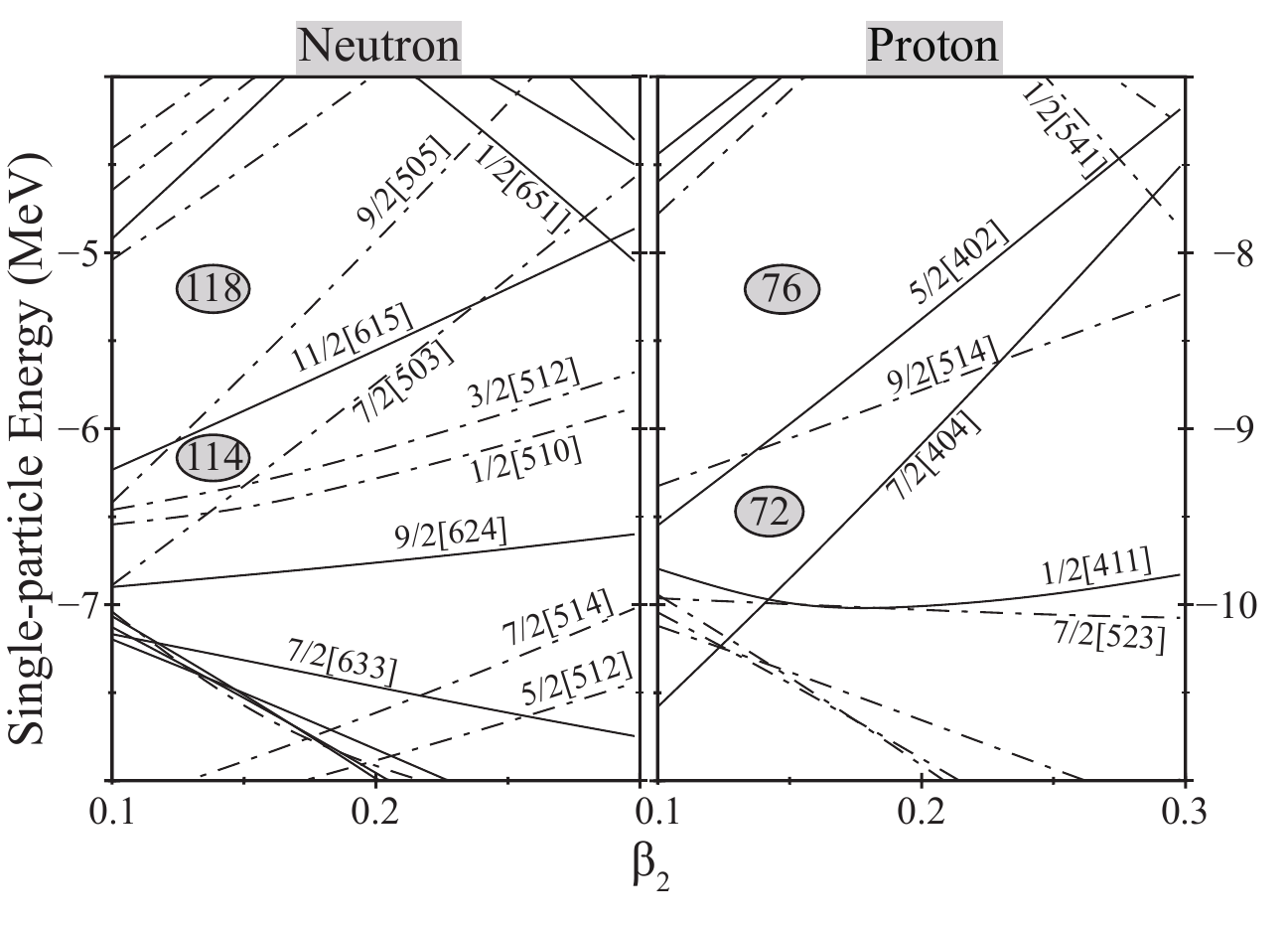}
\caption{Neutron (left) and proton (right) single-particle energies as a function of $\beta_2$-deformation calculated based on a Woods-Saxon potential and universal parameters with $\beta_4 = -0.06$ and $\gamma = 0^{\circ}$ around the $N = 114$ and $Z = 73$ Fermi surfaces.}
    \label{fig:4}
\end{figure}

\begin{table*}[htp!]
\caption{Calculated deformation parameters ($\beta_{2}, \beta_{4}, \gamma$) and excitation energies $E_{\textbf{cal}}$ of selected high-$K$ states with five-quasiparticle configurations in $^{187}$Ta using a configuration-constrained potential-energy surface model.}
\label{tab:2}
\begin{ruledtabular}
\begin{tabular}{llcccc}
$K^{\pi}$ & Configuration &$\beta_{2}$ & $\gamma$ ($^\circ$) & $\beta_{4}$ & $E_{\textbf{cal}}$ (keV) \\
    \hline
$35/2^-$&  ${\pi}\{7/2^+[404]\}\otimes \nu^4\{7/2^-[503], 9/2^-[505], 11/2^+[615], 1/2^-[510]\}$&0.198  &0  &-0.067  &3084  \\
 $37/2^-$&  ${\pi}\{7/2^+[404]\}\otimes \nu^4\{7/2^-[503], 9/2^-[505], 11/2^+[615], 3/2^-[512]\}$&0.198  &-3.5  &-0.067  &2958  \\ 
 $37/2^-$&  ${\pi^3}\{9/2^-[514], 7/2^+[404], 5/2^+[402]\}\otimes \nu^2\{7/2^-[503], 9/2^-[505]\}$&0.195  &-0.3  &-0.065  &3148  \\
 $37/2^+$&  ${\pi}\{9/2^-[514]\}\otimes \nu^4\{7/2^-[503], 9/2^-[505], 11/2^+[615], 1/2^-[510]\}$&0.191  &-0.1  &-0.061  &2967  \\
 $39/2^+$&  ${\pi}\{9/2^-[514]\}\otimes \nu^4\{7/2^-[503], 9/2^-[505], 11/2^+[615], 3/2^-[512]\}$&0.195  &4.8  &-0.061  &2830  \\
 $39/2^+$&  ${\pi^3}\{9/2^-[514], 7/2^+[404], 5/2^+[402]\}\otimes \nu^2\{7/2^-[503], 11/2^+[615]\}$&0.205  &-0.1  &-0.073  &2859  \\
 $41/2^+$&  ${\pi^3}\{9/2^-[514], 7/2^+[404], 5/2^+[402]\}\otimes \nu^2\{9/2^-[505], 11/2^+[615]\}$&0.195  &-0.3  &-0.066  &3152  \\

\end{tabular}
\end{ruledtabular}
\end{table*}

To discuss the structure of $^{187}$Ta$^{m2}$, we have performed configuration-constrained potential-energy-surface (CCPES) calculations~\cite{XU1998257} based on a Woods-Saxon (WS) potential with the so-called universal parameters~\cite{CWIOK1987379}. Figure~\ref{fig:4} shows the SP energies of Nilsson orbits $\Omega^{\pi}[Nn_z\Lambda]$ for neutrons and protons versus the positive (prolate) axial quadrupole deformation parameter $\beta_2$ calculated for $\beta_4 = -0.06$ and $\gamma = 0^{\circ}$ at the ground state of $^{187}$Ta with the same WS potential. Around the $Z=73$ and $N=114$ Fermi surfaces at $\beta_2 \approx 0.2$, three proton ($\pi$) orbitals, $7/2^+[404]$, $9/2^-[514]$, and $5/2^+[402]$, and five neutron ($\nu$) orbitals, $1/2^-[510]$, $3/2^-[512]$, $7/2^-[503]$, $11/2^+[615]$, and $9/2^-[505]$, are considered to contribute to the formation of multi-quasiparticle (qp) states with $K^{\pi}$, which is given as the sum of the respective $\Omega^{\pi}$ values. For instance, the same CCPES calculations indicated that within a narrow range of excitation energy there are two $K^{\pi} = 25/2^-$ states with 3qp configurations of $\pi \otimes \nu^2$ type, $\pi\{ 9/2^-[514]\}\otimes\nu^2\{9/2^-[505], 7/2^-[503]\}$ and $\pi\{7/2^+[404]\}\otimes\nu^2\{11/2^+[615], 7/2^-[503]\} $ (denoted as $[25/2^-]_{\rm A}$ and $[25/2^-]_{\rm B}$, respectively, for later discussion), which are the candidates for $^{187}$Ta$^{m1}$ but unable to be distinguished from an experimental point of view~\cite{PhysRevLett.125.192505}. The largest among the $K$ values of such 3qp states consisting of the aforementioned deformed SP orbitals near the Fermi surfaces is $29/2 \hbar$. Thus, it is necessary to consider 5qp configurations, such as $\pi^3 \otimes \nu^2$ and $\pi \otimes \nu^4$, for the higher-lying isomer $^{187}$Ta$^{m2}$, which can be constrained experimentally to have $K \geq 35/2$, as discussed in the previous paragraph. 

In Table~\ref{tab:2}, calculated excitation energies and deformation parameters ($\beta_2$, $\beta_4$, $\gamma$) for selected 5qp configurations are summarized. These high-$K$ states are predicted to lie around 3 MeV, comparable to the measured excitation energy of $^{187}$Ta$^{m2}$. 
It can be found that, except for the $K^{\pi} = 41/2^+$ case, the configurations listed in Table~\ref{tab:2} include either of the candidate configurations for $^{187}$Ta$^{m1}$, $[25/2^-]_{\rm A}$ or $[25/2^-]_{\rm B}$; thus, they are expected to feed preferentially the levels within the $K^{\pi}=25/2^-$ rotational band at lower excitation energy via $K$-forbidden transitions. For instance, the second $K^{\pi}=39/2^+$ state predicted at 2859 keV is likely to decay towards the $K^{\pi} = 25/2^-$ band member(s) via the $\pi^2\{5/2^+[402],9/2^-[514]\} \rightarrow 0$ ($\mathit{\Delta}K = 7\hbar$) transition involving a parity change with the $[25/2^-]_{\rm B}$ component being a common spectator in the initial and final states. Therefore, in the context of the internal decay patterns, we are unable to make a unique choice in terms of the $K^{\pi}$ and configuration of $^{187}$Ta$^{m2}$ from the calculated levels of $K \geq 35/2$. Despite the difficulty in distinguishing between them, axially symmetric prolate deformation ($\beta_2 \approx 0.2$, $\gamma \approx 0^{\circ}$) is a common feature of the predicted high-$K$ states.

\section{Summary}
\label{sec:summary}
The decay properties of a high-spin, long-lived isomer in $^{187}$Ta, which was previously identified at 2933(14) keV by nuclear mass measurements with the ESR at GSI, have been investigated for the first time using the KEK Isotope Separation System in RIKEN Nishina Center. A half-life of 136(24) s determined in the present paper is much shorter than the value reported at ESR. The internal decay branch was assigned as feeding the $K^{\pi}=(25/2^-)$ isomeric state, while the $\gamma$ rays de-exciting the $K^{\pi}=(11/2^+)$ isomer in the daughter nucleus $^{187}$W have been observed in electron-$\gamma$ delayed coincidence, indicating the occurrence of $\beta$ decay from the high-spin isomer. Based on the evaluated hindrances for $K$-forbidden transitions, $K \geq 35/2$ was proposed for the long-lived isomer in $^{187}$Ta. For these possible $K^{\pi}$ states, configuration-constrained potential-energy-surface calculations predict axially symmetric prolate deformation with five-quasiparticle configurations.

\section*{Acknowledgements}
This experiment was performed at RI Beam Factory operated by RIKEN Nishina Center and CNS, University of Tokyo. The authors thank the RIKEN accelerator staff for their support. This work was funded in part by Grants No. JP23244060, No. JP24740180, No. JP26247044, No. JP15H02096, No. JP17H01132, No. JP17H06090, and No. JP18H03711 from JSPS KAKENHI; No. ST/L005743/1 from United Kingdom STFC; No. 12150710512 from NSFC for International Senior Scientists; No. 11921006, No. 11835001, No. 12335007, No. 12275369 and No. 12035001 from NSFC; No. 682841 “ASTRUm” from ERC (Horizon 2020); and No. DE-AC02-06CH11357 from U.S. Department of Energy (Office of Nuclear Physics).


\nocite{*}
\bibliography{apssamp}
\end{document}